\documentclass[aps,prl,twocolumn,showpacs,preprintnumbers,
amsmath,amssymb]{revtex4}
\usepackage{epsfig}
\usepackage{graphicx}
\usepackage{dcolumn}
\usepackage{bm}

\begin{document}
\title{Influence of high magnetic fields on superconducting transition
of one-dimensional Nb and MoGe nanowires}
\author{A. Rogachev, A.T. Bollinger, and A. Bezryadin}
\affiliation{Department of Physics, University of Illinois at
Urbana-Champaign, Urbana, IL 61801-3080}
\date{\today}

\begin {abstract}
The effects of strong magnetic field on superconducting Nb and
MoGe nanowires with diameter $\sim10$ nm have been studied. We
have found that the Langer-Ambegaokar-McCumber-Halperin (LAMH)
theory of thermally activated phase slips is applicable in a wide
range of magnetic fields and describes well the temperature
dependence of the wire resistance, over eleven orders of
magnitude. The field dependence of the critical temperature,
$T_{c}$, extracted from the LAMH fits is in good quantitative
agreement with the theory of pair-breaking perturbations that
takes into account both spin and orbital contributions. The
extracted spin-orbit scattering time agrees with an estimate
$\tau_{so}\simeq \tau(\hbar c/ Ze^{2})^{4}$, where $\tau$ is the
elastic scattering time and $Z$ is the atomic number.
\end {abstract}


\maketitle


The problem of superconductivity in one-dimensional (1D) systems
attracts much attentions since it involves such fundamental
phenomena as macroscopic quantum tunnelling, quantum phase
transitions and environmental
effects\cite{GiordanoDuanGolubevMatveev,Bezryadin,Lau,CamarotaBuchlerSachdev,
Rogachev,Bollinger,SmithXiong}.
It is expected that a strong magnetic field can be used to control
these phenomena. Indeed, the microscopic theory predicts that a
magnetic field, acting on a superconducting condensate, lifts the
time reversal symmetry of the spin and orbital states of paired
electrons and suppresses the critical temperature, $T_c$
\cite{Suppression,TinkhamBook}. A strong enough field destroys
superconductivity. The magnetic field pair-breaking effects were
studied in depth in two and zero-dimensional systems, i.e. thin
films \cite{Films} and nanograins \cite{Grains}. However, an
experimental verification of the pair breaking effects in 1D
superconductors is long overdue.

A distinct feature of 1D superconductors is the absence of the
phase coherence. Due to fluctuations the amplitude of the order
parameter has a finite probability to reach zero at some point
along the wire, allowing the phase of the order parameter to slip
by $2\pi$ \cite{Little}. The theory of thermally activated phase
slips (TAPS) was developed by Langer, Ambegaokar, McCumber and
Halperin (LAMH). However the effect of the magnetic field on the
phase slippage process is not established. It is also unknown
whether the magnetic field can change the relative contributions
of quantum and thermally activated phase slips in thin wires
\cite{Lau,Rogachev}.

 In this Letter we study the effects of the magnetic
field on the phase slippage rate and the critical temperature of
thin wires. It is found that the LAMH provides a good description
for 1D superconductors in magnetic fields up to ~11 T. The
dependence of the critical temperature on the magnetic field,
$T_c(B)$ agrees well with the theory of pair-breaking
perturbations that takes into account both spin and orbital
contributions \cite{Suppression, TinkhamBook}. This is our main
result. No significant contribution of quantum phase slips has
been detected in the studied samples.

The samples were made by sputter-coating of suspended fluorinated
carbon nanotubes with Nb or Mo$_{79}$Ge$_{21}$. Transport
measurements were performed in a He-3 cryostat, as described
previously \cite{Bezryadin, Rogachev, Bollinger}. The magnetic
field was oriented perpendicular to the wire and parallel to the
thin film electrodes connected in series with the wire.

\begin{figure}[b]
\begin{center}
\epsfig{file=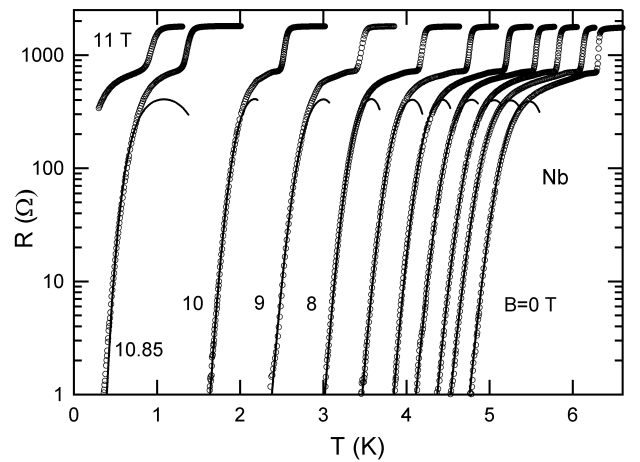, width=3.5 in}\label{fig:NbRT}
\caption{ Resistance versus temperature dependence for a Nb
nanowire (the thickness is 8 nm, the normal resistance is
$R_{N}=700$ $\Omega$, the length is $L$=120 nm). Each R(T) curve
is measured in a fixed magnetic field; some fields are indicated.
Solid lines show the fits to the LAMH theory.}
\end{center}
\end{figure}

A series of resistance versus temperature $R(T)$ curves measured
at different magnetic fields is shown in Fig. 1. For each curve, a
resistance drop at higher temperature corresponds to a
superconducting transition in the film electrodes. The resistance
value immediately below the drop is taken as the wire normal
resistance $R_{N}$. The second resistive transition corresponds to
development of superconductivity in the wire. With increasing
magnetic field, both transitions shift to lower temperatures.

To analyze these data we employ the LAMH expression for zero-bias
resistance:
\begin{equation}
\ R_{LAMH}(T)=\frac{\pi\hbar^{2}\Omega}{2e^{2}kT}e^{-\Delta
F/k_{B}T},
\end{equation}
where $\Delta F=(8\sqrt{2}/3)(H^{2}_{c}/8\pi)A\xi$ is the energy
barrier, $\Omega = (L/\xi )(\Delta F/k_{B}T)^{1/2}(1/\tau _{GL})$
is the attempt frequency, and $\tau_{GL}=[\pi
\hbar/8k_{B}(T_{c}-T)]$ is the Ginzburg-Landau (GL) relaxation
time, $L$ is the length of the wire, $A$ its cross-sectional area,
and $\xi$ is the GL coherence length. Following Ref. \cite{Lau} we
express the energy barrier as
\begin{equation}
\Delta F(T)\approx 0.83 [L/\xi(0)] (R_{q}/R_{N})k_{B}T_{c}
(1-T/T_{c})^{3/2},
\end{equation}
where $R_{q}=h/4e^{2}=6.45$ k$\Omega$ is the resistance quantum
for Cooper pairs and $\xi(0)$ is defined by
$\xi(T)=\xi(0)(1-T/T_{c})^{-1/2}$. Taking into account the
contribution of quasiparticles, the total resistance is given as
$R=(R_{N}^{-1}+R_{LAMH}^{-1})^{-1}$. In our fitting procedure we
use two adjustable parameters, $T_{c}$ and $\xi(0)$.

The LAMH fits are shown as solid lines in Fig. 1, for various
magnetic fields. Although the LAMH theory is derived for B=0, we
find that the resistance agrees with the LAMH fits very well, for
both Nb and MoGe samples, even in high fields up to $\sim$11 T.

Our extension of the LAMH theory to high magnetic fields requires
an explanation: For a phase slip to occur in a wire, the system
needs to overcome an energy barrier that is a product of the
condensation energy density ($H_{c}^{2}/8\pi$) in a volume of a
phase slip $A\xi$. It can be shown within the GL theory that in
magnetic field the condensation energy density goes to zero as
$(1-T/T_{c}(B))^{2}$, where $T_{c}(B)$ is the field-dependent
critical temperature. The coherence length varies as
$\xi(0,B)(1-T/T_{c}(B))^{-1/2}$ and diverges at $T_{c}(B)$.
Because the temperature dependence of both the condensation energy
and the coherence length has the same form as in zero field, we
expect Eq. 1 and Eq. 2 to be applicable in magnetic fields also.
The observed agreement with the experiment suggests that the
mechanism of the phase slippage in 1D wires is not changed by
magnetic field.

To test the LAMH theory in a nonlinear regime, we performed
measurements of the voltage-current $V(I)$ dependencies at high
bias currents for some MoGe wires (Fig. 2a). We observe that at
high currents the wires undergo a transition into resistive state.
This transition is smooth at high temperatures and it is jumpwise,
with some hysteresis, at low temperatures
\cite{Rogachev,Hysteresis} (Fig. 2a). A closer inspection of the
data revealed that even at low temperatures there exists a small
non-linear voltage variation at currents slightly lower than the
switching current (Fig. 2b). The $V(I)$ curves remain
qualitatively unchanged in magnetic field.

\begin{figure}[t]
\begin{center}
\epsfig{file=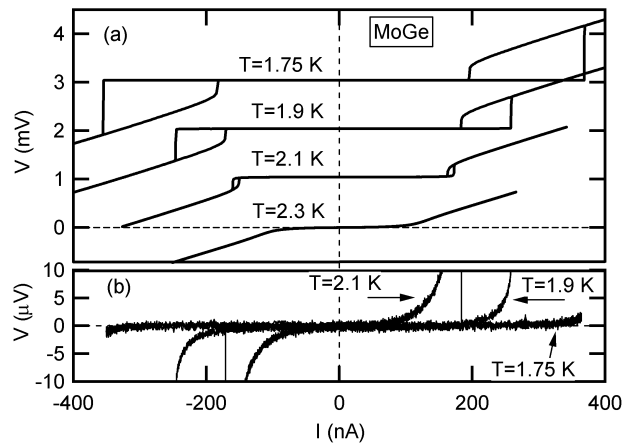, width=3.5 in}
\label{fig:VI}\caption{(a)Voltage versus current dependence at
indicated temperatures for a MoGe nanowire (thickness 7 nm,
$R_{N}=4.3$ k$\Omega$, $L$=190 nm) at $B$=0. The curves are
vertically shifted by 1 mV for clarity. (b) The same data at a
magnified scale.}
\end{center}
\end{figure}

In order to investigate in more detail the resistive tails
observed slightly below the switching current we measured the
differential resistance $dV/dI$ versus bias current $I$ as shown
in Fig. 3.
\begin{figure}[b]
\begin{center}
\epsfig{file=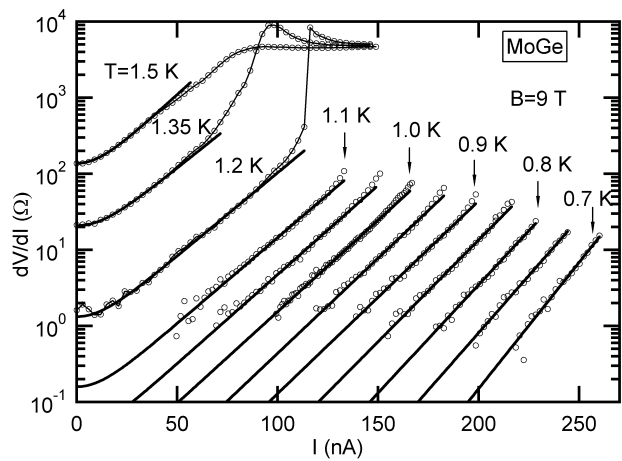, width=3.5 in}
\label{fig:VI}\caption{Differential resistance versus bias current
at indicated temperatures for the MoGe nanowire (thickness 7 nm,
$R_{N}=3.9$ k$\Omega$, $L$=150 nm) in a fixed magnetic field B=9
T. Solid lines are fits to the LAMH expression
$dV/dI=R(T)\cosh(I/I_{0})$}
\end{center}
\end{figure}
It is clear that at all temperatures the $dV/dI$ versus $I$ data
follow exponential dependence, which is expected from the LAMH
expression $dV/dI=R(T)\cosh(I/I_{0})$, where $R(T)$ is the
zero-bias resistance given by Eq. 1 and $I_{0}=4ekT/h$. We fit the
data with the above expression and extract two adjustable
parameters $R(T)$ and $I_{0}$.  Experimental values of $I_{0}$ are
close to the theoretical value $I_{0}=4ekT/h$ (Fig.4, insert). We
speculate that the observed small upward deviation of $I_{0}$
might be due to the fact that some fraction of the bias current is
carried by nonequilibrium quasiparticles. Such quasiparticles are
generated by the phase slips and are not taken into account within
the LAMH. A detailed theoretical analysis is needed for further
understanding of the $I_0(T)$ behavior.

In Fig. 4 we superimpose the zero-bias resistance data and the
resistance data obtained from the fits to the nonlinear portion of
$dV/dI$ vs. $I$ curves. These two sets of data appear mutually
consistent. The LAMH fit (solid lines in Fig. 4) gives an
excellent description of all resistance data in a range of
\textit{eleven orders of magnitude}. We also find a good agreement
with the LAMH for data taken in magnetic field $B$=9 T (Fig. 4).
It is therefore concluded that the resistance in studied nanowires
is determined by thermally activated phase slips even in high
magnetic fields.

\begin{figure}[t]
\begin{center}
\epsfig{file=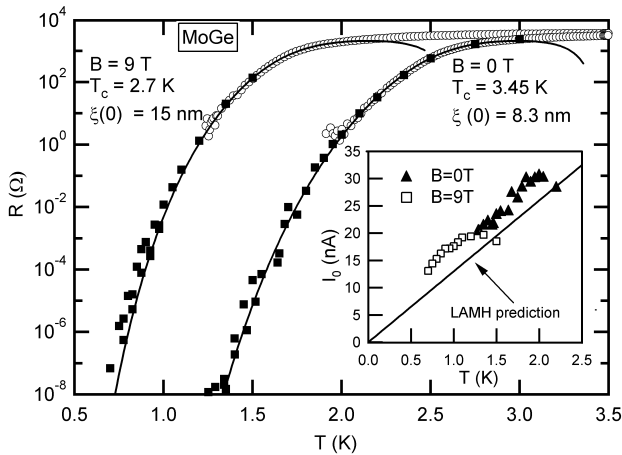, width=3.5 in}
\label{fig:RT_nonlinear}\caption{Resistance versus temperature for
the MoGe nanowire (same as in Fig. 3) in magnetic fields 0 and 9
T. Open circles represent zero-bias-current measurements and black
squares indicate the resistance values obtained from the fit of
the nonlinear portion of $dV/dI$ curves (Fig. 3). Solid lines are
the fits to the LAMH theory (Eq.1). Extracted fitting parameters
$T_{c}$ and $\xi(0)$ are indicated. The insert shows the
experimental dependence of the parameter $I_{0}$ on temperature
(solid and open symbols) and the theoretical value $I_{0}=4ekT/h$
(solid line).}
\end{center}
\end{figure}

\begin{figure}[t]
\begin{center}
\epsfig{file=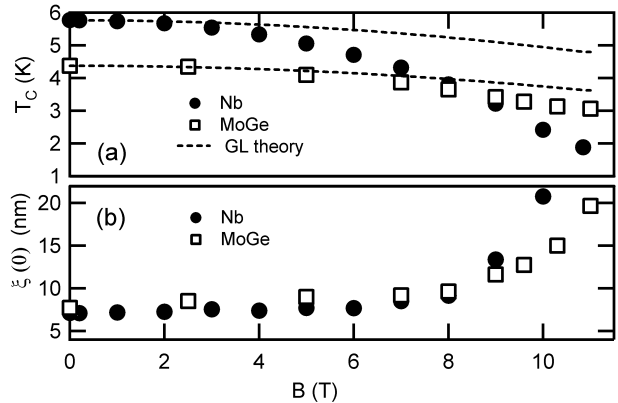, width=3.5 in}
\label{fig:TcXi}\caption{Adjustable parameters used in Eq.1 for
the LAMH fits of the type shown in Fig.1. (a) The critical
temperature versus magnetic field $T_{c}(B)$ is shown for the Nb
wire (same as in Fig.1) and for a MoGe wire. (b) The GL coherence
length versus magnetic field.  The parameters of the MoGe wire
are: thickness is 9 nm, $R_{N}=7.5$ k$\Omega$, $L$=460 nm. The
theoretical Ginzburg-Landau dependence is indicated by dashed
lines.}
\end{center}
\end{figure}

In Fig. 5 we plot the extracted parameters $T_c(B)$ and $\xi(0,B)$
versus magnetic field. The coherence length increases very slowly
in low fields and starts to grow more rapidly at $B>8$ T. We
consider now the critical temperature $T_c(B)$, which is the main
focus of this work. Although the initial decrease of $T_{c}$
agrees with the variation predicted by the GL theory (Fig. 5a), it
is clear that the GL phenomenology is not sufficient to account
for the observed $T_c(B)$ dependence. One possible reason is that
the GL theory does not take into account the spin pair-breaking
effect and the coefficients of the GL theory can change in the
high field regime. Thus we have to use the exact theory of
pair-breaking \cite{Suppression}. In both Nb and MoGe nanowires
superconductivity persists to magnetic fields that are larger than
the paramagnetic limit, $B_{p}$[T]=1.84 $T_{c}$ [K]
\cite{TinkhamBook}, (10.6 T for Nb and 8.1 T for MoGe). Since the
superconductivity is not fully suppressed at such fields, we
conclude that the effect of magnetic field on the spin part of the
Cooper pair is reduced by the spin-orbit scattering, as expected
for materials with high atomic numbers. If the spin-orbit
scattering is sufficiently strong, the transition into the normal
sate is continuous. This allows us to take into account both the
spin and the orbital effects on $T_{c}$ in an implicit relation of
the theory of pair-breaking perturbations \cite{TinkhamBook}

\begin{equation}
\ln\frac{T_{c}(B)}{T_{c}(0)}=\psi \left( \frac{1}{2} \right)-\psi
\left(\frac{1}{2}+\frac{\alpha_{o}+\alpha_{s}}{2\pi k_B
T_{c}(B)}\right),
\end{equation}

where $\psi(z)$ is the digamma function and
$\alpha_{o}=2De^{2}\langle A^{2} \rangle/\hbar c^{2}$ and
$\alpha_{s}\approx\tau_{so} e^{2} \hbar B^{2}/m^{2}c^{2}$ are the
orbital and the spin pair-breaker strength parameters, $D$ is the
diffusion coefficient and $\tau_{so}$ is the spin-orbit scattering
time. To find  the vector potential averaged over the
cross-sectional area of a wire, $\langle A^{2} \rangle$, we use
the expression for a cylinder in perpendicular magnetic field,
$\langle A^{2} \rangle=B^{2}d^{2}/16$, where $d$ is the diameter
of the wire.

By solving Eq. 3 numerically the theoretical dependence of the
normalized critical temperature versus the normalized magnetic
field is obtained. In Fig.6 this dependence is compared to the
experimental values of $T_c(B)/T_c(0)$ (same data as in Fig.5a).
Since both pair-breakers, $\alpha_{o}$ and $\alpha_{s}$, have a
quadratic dependence on $B$, only one adjustable parameter is
required, which is the critical field of the wire at zero
temperature, $B_{wc}$. The best fits are obtained by choosing
$B_{wc}$=11.7 T for Nb and 16.5 T for MoGe (Fig. 6).

\begin{figure}[t]
\begin{center}
\epsfig{file=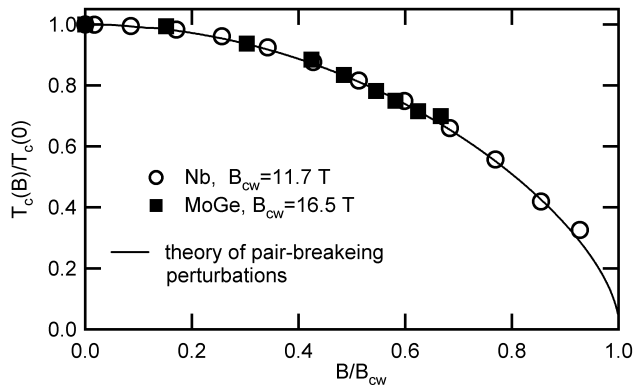, width=3.5 in}
\label{fig:G_Tc_p}\caption{Normalized critical temperature versus
normalized magnetic field for the Nb and MoGe wires. Critical
magnetic fields ($B_{cw}$) for both wires are indicated. The solid
line is a fit to the theory of pair-breaking perturbation (Eq.
4).}
\end{center}
\end{figure}

To compare orbital and spin pair-breaking contributions we
introduce an orbital critical field ($B_{co}$) and a spin critical
field ($B_{cs}$), defined as fields needed to suppress
superconductivity if only one of the two pair-breaking mechanisms
is present. These fields are obtained from equations
$2\alpha_{o}=2\alpha_{s}=1.76kT_{c}(0)$ \cite{TinkhamBook}. The
total critical field is then
$B_{wc}=(B_{co}^{-2}+B_{cs}^{-2})^{-1/2}$. Using BCS and GL
relations the orbital critical field can be written as
$B_{co}=0.53\Phi_{0}/d\xi(0)$, where $\Phi_{0}$ is the flux
quantum. The value of the coherence length $\xi(0)$ is known from
the LAMH fit in zero field (7.1 nm for Nb and 7.7 nm for MoGe).
Taking as $d$ the thickness of the deposited material (corrected
for oxidation for the MoGe wire \cite{Bollinger}) we estimate
$B_{co}\approx22$ T for both wires. From $B_{co}$ and experimental
total critical field $B_{wc}$ we determine the spin critical
fields, $B_{cs}\approx14$ T for Nb and $B_{cs}\approx25$ T for
MoGe. Thus the spin and orbital pair-breakers have comparable
strengths.

From the $B_{cs}$ we estimate the spin-orbit scattering time as
$\tau_{so}=2.3\pm 0.5\times 10^{-13}$ s for the Nb wire and
$\tau_{so}=5\pm 3\times 10^{-14}$ s for the MoGe wire. The MoGe
result is considerably different from the value
$\tau_{so}\simeq1.3\times10^{-12}$ s obtained for thin MoGe films
from weak localization measurements \cite{Graybeal}. With such
$\tau_{so}$ value the superconductivity in MoGe wire would be
completely suppressed at $B_{cs}\simeq 5$ T, contrary to our
observation. On the other hand, we can use the formula
$\tau_{so}\simeq \tau(\hbar c/ Ze^{2})^{4}$ given in Ref.
\cite{Abrikosov}. With elastic scattering time $\tau\simeq
6\times10^{-16}$ s \cite{Graybeal} this gives a shorter spin-orbit
scattering time $8\times10^{-14}$ s, that agrees with our result.
The latter estimate also works well for the Nb nanowire. Here we
have elastic time $\tau=\ell/v_{F} \simeq 1.9\times10^{-15}$ s
($v_{F}=0.62\times10^{-8}$ cm/sec \cite{Mattheiss}, $\ell$=1.2 nm
\cite{Rogachev}), and so $\tau_{so} \simeq  2.4 \times10^{-13}$ s,
again in agreement with the experimental value. Thus we conclude
that the pair-breaking theory combining spin and orbital
contributions gives an accurate quantitative prediction for the
suppression of the critical temperature of homogeneous 1D
superconducting wires.

This work is supported by NSF CAREER Grant No. DMR 01-34770, by
A.P.Sloan Foundation and by a DOE grants DEFG02-91ER45439 and
DEFG02-96ER45439.


\end{document}